\documentclass
[superscriptaddress,secnumarabic,amssymb,amsmath,nobibnotes,aps,prd,showkeys,showpacs,nofootinbib,onecolumn]{revtex4}%
\usepackage{graphicx}
\usepackage{bm}
\usepackage{bigints}
\usepackage{amsmath}
\usepackage{amsfonts}
\usepackage{amssymb}%
\providecommand{\U}[1]{\protect\rule{.1in}{.1in}}

\newcommand{\be}{\begin{equation}}
\newcommand{\ee}{\end{equation}}

\newcommand{\mincir}{\raise
-3.truept\hbox{\rlap{\hbox{$\sim$}}\raise4.truept\hbox{$<$}\ }}
\newcommand{\magcir}{\raise
-3.truept\hbox{\rlap{\hbox{$\sim$}}\raise4.truept\hbox{$>$}\ }}

\begin{document}
\title{Analytic solutions in Einstein-aether scalar field cosmology}
\author{Andronikos Paliathanasis}
\email{anpaliat@phys.uoa.gr}
\affiliation{Institute of Systems Science, Durban University of Technology, Durban 4000,
South Africa}
\author{Genly Leon}
\email{genly.leon@ucn.cl}
\affiliation{Departamento de Matem\'{a}ticas, Universidad Cat\'{o}lica del Norte, Avda.
Angamos 0610, Casilla 1280 Antofagasta, Chile.}

\begin{abstract}
In the context of Einstein-aether scalar field cosmology we solve the field equations and determine exact and analytic solutions. In particular, we consider a model proposed by Kanno and Soda where the aether and the scalar fields interact through the aether coefficient parameters, which are promoted to be functions of the scalar field. For this model, we write the field equations by using the minisuperspace approach and we determine the scalar field potentials which leads to Liouville--integrable systems. We solve the field equations for five families of scalar field potentials  and, whether it is feasible, we write down the analytic solutions by using closed-form functions. 
\end{abstract}
\date{\today}
\maketitle

\section{Introduction}

According to well verified observations the universe is currently exhibiting
an accelerated expansion. Although the simplest explanation would be the
cosmological constant \cite{Weinberg:1988cp}, the possible dynamical features
require for more radical modifications. Hence, one can either alter the
universe content, by introducing new, exotic forms, collectively called
\textquotedblleft dark energy\textquotedblright%
\ \cite{Copeland:2006wr,Cai:2009zp}, e.g., quintom models \cite{Cai:2009zp,Dutta:2009yb,Guo:2004fq,Zhao:2006mp,Lazkoz:2006pa,Lazkoz:2007mx,MohseniSadjadi:2006hb,Setare:2008pz,Setare:2008dw,Saridakis:2009ej,Qiu:2010ux,Leon:2012vt,Leon:2018lnd,Paliathanasis:2018vru}, which generalizes phantom fields \cite{Singh:2003vx,Sami:2003xv,Andrianov:2005tm,Elizalde:2008yf,Sadatian:2008sv}, or modify the gravitational sector adding
new degrees of freedom, like in $f$- theories,
\cite{DeFelice:2010aj,Nojiri:2005jg}, in Lovelock gravity
\cite{Lovelock:1971yv,Deruelle:1989fj}, in scalar field theories like the
Galileon theory
\cite{Nicolis:2008in,Deffayet:2009mn,Leon:2012mt,DeArcia:2015ztd,Dimakis:2017kwx,Giacomini:2017yuk,DeArcia:2018pjp}%
; or in the Lorentz invariant Ho\v{r}ava-Lifshitz gravity
\cite{Horava:2009uw,Leon:2009rc,Leon:2019mbo},  and many others.

A very interesting theory of gravitational modification is the Einstein-aether
theory
\cite{Jacobson:2000xp,Eling:2004dk,Carroll:2004ai,Kanno:2006ty,Zlosnik:2006zu,Donnelly:2010cr,Carruthers:2010ii,Jacobson:2010mx,Jacobson,Garfinkle:2011iw,Barrow:2012qy,Sandin:2012gq,Alhulaimi:2013sha,Coley:2015qqa,Latta:2016jix,Alhulaimi:2017ocb,VanDenHoogen:2018anx,Coley:2019tyx,Leon:2019jnu}%
. It corresponds to the class of Lorentz-violating theories of gravity, where
one considers the existence of a unit vector, the aether, which is everywhere
non-zero in any solution. The aether spontaneously breaks the boost sector of
the Lorentz symmetry by selecting a preferred frame at each point in
spacetime while maintaining local rotational symmetry. The action for
Einstein-aether theory is the most general generally covariant functional of
the spacetime metric $g_{ab}$ and aether field $u^{a}$ involving no more than
two derivatives, excluding total derivatives
\cite{Jacobson,Carroll:2004ai,Garfinkle:2011iw}. Exact solutions and
qualitative analysis of Einstein-aether were presented elsewhere, e.g., in
\cite{Barrow:2012qy,Sandin:2012gq,Alhulaimi:2013sha,Coley:2015qqa,Latta:2016jix,Alhulaimi:2017ocb,VanDenHoogen:2018anx,Coley:2019tyx,Leon:2019jnu}%
.

In \cite{Kanno:2006ty} it was explored the impact of Lorentz violation on the
inflationary scenario. More precisely, it is studied homogeneous but
anisotropic solutions in the presence of a positive cosmological constant,
with a Bianchi type I (Kasner-like) symmetry with three orthogonal principal
directions of expansion, and with the aether tilted in one of the principal
directions. In this model the inflationary stage is divided into two parts;
the Lorentz violating stage and the standard slow-roll stage. In the first
stage the universe expands as an exact de Sitter spacetime, although the
inflaton field is rolling down the potential. Interestingly, exact Lorentz
violating inflationary solutions can be found in the absence of an inflaton
potential. To linear order in the anisotropy, the system relaxes exponentially
to the isotropic, de Sitter solution. This approach was an special case of the
perturbative treatments used in \cite{Carruthers:2010ii}. In
\cite{Carruthers:2010ii}, it was investigated large deviations from isotropy,
maintaining homogeneity. It was found that, for generic values of the coupling
constants, the aether and metric isotropizes if the initial aether hyperbolic
boost angle and its time derivative in units of the cosmological constant are
less than order $\mathcal{O}(1)$. For larger angles or larger angle derivatives, the
behavior is strongly dependent on the values of the coupling constants.
In general, there is a runaway behavior in which the anisotropy increases with
time, and/or singularities occur. In \cite{Donnelly:2010cr} it was studied the
Einstein-aether theory with an scalar inflaton coupled bilinearly to the
expansion of the aether. There were determined the
conditions for linearized stability, positive energy, and vanishing of
preferred-frame post-Newtonian parameters, and examined whether all of these 
restrictions can be simultaneously satisfied. In a homogeneous and isotropic
cosmology, the inflaton-aether expansion coupling leads to a driving force on
the inflaton that is proportional to the Hubble parameter. This force affects
the slow-roll dynamics, but still allows a graceful exit of inflation.

Einstein-aether theory have been applied also in various anisotropic and
inhomogeneous models with many interesting results. In \cite{Coley:2015qqa}
were studied spherically symmetric cosmological models in Einstein-aether
theory with a non-comoving perfect fluid source using a 1+3 frame formalism,
in the context of inhomogeneous cosmological models. Adopting the comoving
aether gauge it is derived the evolution equations in normalized variables to
provide numerical computations and studying the local stability of the
equilibrium points of the resulting dynamical system. Special emphasis was
made on spatially homogeneous Kantowski-Sachs models, see also
\cite{Latta:2016jix,Alhulaimi:2017ocb,VanDenHoogen:2018anx}. 
In \cite{Alhulaimi:2017ocb} was studied the dynamics of spatially homogeneous (SH)
Einstein-aether cosmological models with an scalar field with a self-interaction generalized
harmonic potential, in which the scalar field is coupled to both the aether field
expansion and shear scalars. The stability analysis indicated that there exists a range of values of the
parameters where the late-time attractor corresponds to an accelerated
expansion phase. For the analysis are considered spatially curvature and
anisotropic perturbations. On the other hand, static anisotropic
models for a mixture of a necessarily non-tilted perfect fluid with a
barotropic equation of state (linear and/or polytropic equations of state) and a self-interacting scalar field were studied in
\cite{Coley:2015qqa,Coley:2019tyx,Leon:2019jnu}.  In \cite{Roumeliotis:2019tvu} it was presented the
solution space of the field equations in the Einstein-aether theory for the
case of a vacuum Bianchi Type V space-time. In this model the reduced
equations not always admits a solution. Whenever a solution do exist, their
physical interpretation was examined through the analysis of the behavior of Ricci and/or
Kretschmann scalar, as well as with the identification of the effective energy
momentum tensor in terms of a perfect fluid. There are cases in which no
singularities appears and in other cases the effective fluid is isotropic. Friedmann--Lema\^{\i}tre--Robertson--Walker metric (FLRW) and
a Locally Rotationally Symmetric (LRS) Bianchi Type III space-time were studied in \cite{Roumeliotis:2018ook}.
It was examined whether the reduced equations do have a solution,  and it was found that
there are portions of the initial parameters space for which no solution is
admitted by the reduced equations.

In \cite{Paliathanasis:2020bgs} it is considered an Einstein-aether scalar
field cosmological model where the aether and the scalar field are
interacting through two different interactions proposed in the literature by
Kanno and Soda \cite{Kanno:2006ty} and by Donnelly et al. \cite{Donnelly:2010cr}%
. It was provided an extended dynamical systems analysis of the cosmological
evolution. The reduced Lagrangians deduced from the full action are, in
general, correctly describing the dynamics whenever solutions do exist.
Furthermore, the cosmological evolution of the field equations in the context
of Einstein-aether cosmology by including a scalar field in a spatially flat
FLRW spacetime was studied in \cite{Paliathanasis:2019pcl} by using dynamical system tools. The
analysis was separated into two cases: a pressureless fluid source is included
or it is absent. The limit of general relativity is fully recovered, while the
dynamical system admits de Sitter solutions which can describe the past
inflationary era and the future late-time attractor. Results for generic
scalar field potentials were presented, while some numerical behaviors were
given for specific potential forms. 

The plan of the paper is as follows. 

In Section \ref{sec2}, we present the cosmological model under consideration
which is that of the Einstein-aether gravity with an scalar field coupled to the aether through an effective coupling $B\left(  \phi\right)  $ as defined by $B\left(  \phi\right)
=\beta_{1}\left(  \phi\right)  +3\beta_{2}\left(  \phi\right)  +\beta
_{3}\left(  \phi\right)  -1$ in terms of the aether parameters $\beta_1, \ldots \beta_4$ 
\cite{Kanno:2006ty}. As far as for the physical space is concerned, we consider
that it is described by the spatially flat FLRW metric. For the latter
cosmological model we present the field equations and we give the
minisuperspace description of the theory as well. 

In Section \ref{sec3}, we determine exact solutions of the field equations of physical interest.
Specifically we find the scalar field potential such
that the scale factor of the FLRW metric describes either the de Sitter universe, or
it describes an scaling solution. In addition, we study the stability of these
solutions by calculating their first order perturbations around the
exact solutions and analyzing their evolution. 

The main results of our analysis are presented in Section \ref{sec4}. 
We assume the presence of a dust fluid in the cosmological model. We determine
the functional forms of the scalar field potential such that the field
equations are Liouville--integrable, with at least the existence of a second
conservation law, quadratic in the momentum. We find five families of power-law potentials, 
for which we present the analytic solutions as functions
in a close-form or in algebraic form by solving the Hamilton-Jacobi equations
and reducing the dimensionality of the field equations. By using the results
of Section \ref{sec3}, we infer the asymptotic behavior of the cosmological
solutions, since we can relate the dominant terms of the scalar field
potential with the exact solutions presented in Section \ref{sec3}. Recall that a system of polynomial differential equations is said to be Liouville--integrable, if it has first integrals given by elementary functions or integrals of elementary functions, that is, functions expressed in terms of combinations of exponential functions, trigonometric functions, logarithmic functions or polynomial functions (see, e.g., \cite{Iacono:2014uga}, in the context of Tolman-Oppenheimer-Volkoff approach for a relativistic star model
with the isothermal equation of state $p_{m}=\rho_{m}/n$; which is Liouville--integrable if and only if $n\in\{-1, -3, -5, -6\}$). 

In Appendix \ref{appa}, we present the five Liouville integrable scalar field
potentials where the additional matter source in the cosmological fluid it is
an ideal gas with equation of state $p_{m}=\left(  \gamma-1\right)  \rho_{m}$.
Note that when $\gamma=\frac{2}{3}$, our results describe the case of a non
spatially flat FLRW spacetime. 

Finally, in Section \ref{sec5}, we summarize
the results and we draw our conclusions.

\section{Einstein-aether Scalar field Cosmology}

\label{sec2}

We consider the Einstein-aether scalar field theory with Action Integral \cite{Kanno:2006ty}:
\begin{equation}
S=\int dx^{4}\sqrt{-g}\left(  \frac{R}{2}-\frac{1}{2}g^{\mu\nu}\phi_{;\mu}%
\phi_{;\nu}-V\left(  \phi\right)  \right)  -S_{\text{Aether}}, \label{ac.01}%
\end{equation}
where $S_{\text{Aether}}$ describes the terms of the aether field $u^{\mu}$ as
follows
\begin{align}
S_{\text{Aether}}  &  =\int dx^{4}\sqrt{-g}\left(  \beta_{1}\left(  \phi\right)
u^{\nu;\mu}u_{\nu;\mu}+\beta_{2}\left(  \phi\right)  u^{\nu;\mu}u_{\mu;\nu
}\right)  +\nonumber\\
&  +\int dx^{4}\sqrt{-g}\left(  \beta_{3}\left(  \phi\right)  \left(
g^{\mu\nu}u_{\mu;\nu}\right)  ^{2}+\beta_{4}\left(  \phi\right)  u^{\mu}%
u^{\nu}u_{;\mu}u_{\nu}-\lambda\left(  u^{\mu}u_{\nu}+1\right)  \right)  .
\label{ac.02}%
\end{align}
Coefficients $\beta_{1},~\beta_{2},~\beta_{3}~$and $\beta_{4}$ define the
coupling between the aether field and the gravitational field. In
Einstein-aether theory, the coefficients are constants, though in this model,
coefficients $\beta_{1},~\beta_{2},~\beta_{3}~$and $\beta_{4}$ define a coupling
between the aether field $u^{\mu}$ and the scalar field $\phi\left(  x^{\mu
}\right)$, by promoting themselves to be functions of $\phi$. Additionally, function $\lambda$ is a Lagrange multiplier which
ensures the unitarity, $u^{\mu}u_{\mu}+1=0$, of the aether field $u^{\mu}$.

In large scales the universe it is assumed to be isotropic and homogeneous
described by the spatially flat FLRW metric, with line element%
\begin{equation}
ds^{2}=-N^{2}\left(  t\right)  dt^{2}+a^{2}\left(  t\right)  \left(
dx^{2}+dy^{2}+dz^{2}\right)  , \label{ac.03}%
\end{equation}
where $a\left(  t\right)  $ is the scale factor, $N\left(  t\right)  $ is the
lapse function while the Hubble function is defined as $H\left(  t\right)
=\frac{1}{N}\frac{\dot{a}}{a}$, where a dot denotes total derivative with
respect the variable $t$.

For the aether field $u^{\mu}=\frac{1}{N}\delta_{t}^{\mu}$, and the line
element (\ref{ac.03}), the Action Integral (\ref{ac.01}) is simplified as
follows \cite{Kanno:2006ty}:%
\begin{equation}
S=\int dx^{4}\sqrt{-g}L\left(  N,a,\dot{a},\phi,\dot{\phi}\right),
\label{ac.04}%
\end{equation}
where~$L\left(  N,a,\dot{a},\phi,\dot{\phi}\right)  $ is the point-like
Lagrangian
\begin{equation}
L\left(  N,a,\dot{a},\phi,\dot{\phi}\right)  =\frac{1}{N}\left(  -3B\left(
\phi\right)  a\dot{a}^{2}+\frac{1}{2}a^{3}\dot{\phi}^{2}-N^{2}a^{3}V\left(
\phi\right)  \right)  , \label{ac.05}%
\end{equation}
while function $B\left(  \phi\right)  $ is defined as $B\left(  \phi\right)
=\beta_{1}\left(  \phi\right)  +3\beta_{2}\left(  \phi\right)  +\beta
_{3}\left(  \phi\right)  -1$, and we have assumed that the scalar field $\phi$
inherits the symmetries of the spacetime such that $\phi=\phi\left(  t\right)
$.

Variation with respect the variables $a$ and $\phi$ of the Action Integral
(\ref{ac.04}) gives the second-order field equations%
\begin{equation}
\left(  2\ddot{a}-\frac{2}{N}a\dot{a}\dot{N}\right)  B\left(  \phi\right)
+2aB_{,\phi}\dot{a}\dot{\phi}+B\left(  \phi\right)  \dot{a}^{2}+\frac{1}%
{2}a^{2}\dot{\phi}^{2}-N^{2}a^{2}V\left(  \phi\right)  =0, \label{ac.06}%
\end{equation}%
\begin{equation}
\ddot{\phi}+3\frac{\dot{a}}{a}\dot{\phi}-\frac{\dot{N}}{N}\dot{\phi}+\frac
{3}{A^{2}}B_{,\phi}\dot{a}^{2}+N^{2}V_{,\phi}=0. \label{ac.07}%
\end{equation}
Equation (\ref{ac.07}) is the modified Klein-Gordon equation for the scalar
field $\phi$, while equation (\ref{ac.06}) is the modified second Friedmann
equation. Moreover, variation of (\ref{ac.04}) with respect to the variable $N$
produces the modified first Friedmann equation, that is, the constraint
equation,%
\begin{equation}
-\frac{3}{N^{2}}B\left(  \phi\right)  a\dot{a}^{2}+\frac{1}{2N^{2}}a^{3}%
\dot{\phi}^{2}+a^{3}V\left(  \phi\right)  =0. \label{ac.08}%
\end{equation}

The field equations (\ref{ac.06}), (\ref{ac.07}) and (\ref{ac.08}) can be
written as follows%
\begin{equation}
3H^{2}=k_\text{eff}\rho_\text{eff}, \label{ac.09}%
\end{equation}%
\begin{equation}
-\left(  2\dot{H}+3H^{2}\right)  =k_\text{eff}p_\text{eff}, \label{ac.10}%
\end{equation}
and%
\begin{equation}
\ddot{\phi}+3H\dot{\phi}+3B_{,\phi}H^{2}+V_{,\phi}=0, \label{ac.11}%
\end{equation}
where $k_\text{eff}=\frac{1}{B\left(  \phi\right)  }$, and $\rho_\text{eff}$ and
$p_\text{eff}$ describe the energy density and the pressure of the effective fluid,
defined as
\begin{align}
\rho_\text{eff}  &  =\frac{1}{2}\dot{\phi}^{2}+V\left(  \phi\right),
\label{ac.12}\\
p_\text{eff}  &  =\left(  2B_{,\phi}H\dot{\phi}+\frac{1}{2}\dot{\phi}^{2}-V\left(
\phi\right)  \right). \label{ac.13}%
\end{align}

We observe that there are similarities with the Scalar-tensor theories as
mentioned in \cite{Kanno:2006ty}, indeed $k_\text{eff}$ is not a constant but
changes in time, however the two theories are different. The effective
$k_\text{eff}$ is the only contribution of the aether field in the first Friedmann
equation, because the effective energy density $\rho_\text{eff}$ is that of the
scalar field, i.e. $\rho_\text{eff}=\rho_{\phi}$. On the other, hand from second
Friedmann equation we see that the $2B_{,\phi}H\dot{\phi}$ \ modifies the
effective pressure from that of the scalar field, that is, $p_\text{eff}%
=p_{\phi}+2B_{,\phi}H\dot{\phi}$. Consequently, the parameter for the
effective equation of state it is defined as
\begin{equation}
w_\text{eff}=w_{\phi}+\frac{2B_{,\phi}H\dot{\phi}}{\rho_\text{eff}}. \label{ac.14}%
\end{equation}

As far as equation (\ref{ac.11}) is concerned, this reads%
\begin{equation}
\dot{\rho}_{\phi}+3H\left(  \rho_{\phi}+p_{\phi}\right)  =-3B_{,\phi}%
H^{2}.\label{ac.15}%
\end{equation}
which looks like the particle creation, bulk viscosity or varying vacuum
theories \cite{par1,par2,par3,par4,par5,par6,par7}. Positive values of
$B_{\phi}$ indicate particle annihilation while negative values of $B_{\phi}$
indicate particle creation.

In the presence of an additional fluid source, such that of an ideal gas
$p_{m}=\left(  \gamma-1\right)  \rho_{m}$ which we assume that it is not
interacting with the scalar field or with the aether field, the effective energy
density and pressure terms are modified as
\begin{align}
\rho_\text{eff}  &  =\frac{1}{2}\dot{\phi}^{2}+V\left(  \phi\right)  +\rho
_{m},\label{ac.16}\\
p_\text{eff}  &  =\left(  2B_{,\phi}H\dot{\phi}+\frac{1}{2}\dot{\phi}^{2}-V\left(
\phi\right)  \right)  +\left(  \gamma-1\right)  p_{m}, \label{ac.17}%
\end{align}
with the additional conservation equation%
\begin{equation}
\dot{\rho}_{m}+3\gamma H\rho_{m}=0, \label{ac.18}%
\end{equation}
from which we infer $\rho_{m}=\rho_{m0}a^{-3\gamma}$, $\rho_{m0}$ is an
integration constant. In the following Section, we assume that the additional
matter source is that of a dust fluid, that is, $\gamma=1$ and $\rho_{m}%
=\rho_{m0}a^{-3}$.

\section{Exact solutions}

\label{sec3}

In this section we present some exact solutions of the field equations. 
In particular we determine the functional forms of the potential $V\left(
\phi\right)$ and the function $\phi\left(t\right)$ by incorporating the 
requirement that the de Sitter solution $a\left(t\right)  =a_{0}e^{H_{C}t}$ 
and the scaling solution $a\left(  t\right)  =a_{0}t^{p}$, are special solutions 
of the field equations. 

Recall that we have assumed that $N\left(t\right)=1$. In addition we
assume that there is not any contribution in the cosmological fluid by the
ideal gas, i.e. $\rho_{m0}=0$. Because there are only two independent equations and there are three unknown
functions, namely, $\phi\left(  t\right),~V\left(  t\right)$ and $B\left(t\right)$, we proceed further by 
defining the exact form of $B\left(\phi\left(t\right)\right)$. In particular we select $B\left(\phi\left(t\right)\right)=6B_{0}\phi^{2}$.

\subsection{De\ Sitter solution}

The exponential scale factor $a\left(  t\right)  =a_{0}e^{H_{0}t}$ solves the
field equations (\ref{ac.06})-(\ref{ac.08}) if and only if \cite{Kanno:2006ty}%
\begin{equation}
\phi\left(  t\right)  =\phi_{0}e^{-24\sqrt{B_{0}}H_{0}t}~,~V\left(  t\right)
=18\left(  1-16B_{0}\right)  \phi_{0}^{2}H_{0}^{2}e^{-48H_{0}t},
\end{equation}
that is%
\begin{equation}
V\left(  \phi\left(  t\right)  \right)  =18\left(  1-16B_{0}\right)  H_{0}%
^{2}\phi^{2}.
\end{equation}

In order to determine the stability of the de Sitter solution we substitute
$a\left(  t\right)  =a_{0}e^{\frac{H_{0}}{\sqrt{B_{0}}}t}+\varepsilon\delta
a\left(  t\right)  $ and $\phi\left(  t\right)  =\phi_{0}e^{-24\sqrt{B_{0}%
}H_{0}t}+\varepsilon\delta\phi\left(  t\right)  $ in the field equations and
we linearize around $\varepsilon\rightarrow0$. We end with the linearized
system%
\begin{align}
\delta\ddot{a}+\left(  \frac{1}{B_{0}}-48\right)  \left(  \sqrt{B_{0}}%
\delta\dot{a}-2H_{0}\delta\dot{a}^{2}\right)   &  =0,\\
\delta\ddot{\phi}+3\frac{H_{0}}{\sqrt{B_{0}}}\dot{\phi}+72\left(
1-8B_{0}\right)  H_{0}^{2}\delta\phi &  =0,
\end{align}
with closed-form solution%
\begin{align}
\delta a  &  =\delta a_{0}e^{\frac{H_{0}}{\sqrt{B_{0}}}t}+\delta
a_{1}e^{-\frac{2}{\sqrt{B_{0}}}\left(  1-24B_{0}\right)  H_{0}t},\\
\delta\phi &  =\delta\phi_{0}e^{-24\sqrt{B_{0}}H_{0}t}+\delta\phi_{1}%
e^{-\frac{3}{\sqrt{B_{0}}}\left(  1-8B_{0}\right)  H_{0}t}.
\end{align}
Therefore we conclude that the expanding de Sitter universe is stable when
$0<B_{0}<\frac{1}{24}$, while when $H_{0}<0$ the exact solution is stable when
$B_{0}>\frac{1}{24}$.

\subsection{Scaling solution}

In a similar way, we find that the scaling solution $a\left(  t\right)
=a_{0}t^{\frac{p}{\sqrt{B_{0}}}}$ satisfies the field equations (\ref{ac.06}%
)-(\ref{ac.08}) when \cite{Kanno:2006ty}%
\begin{equation}
\phi\left(  t\right)  =\phi_{0}t^{-12p-2\sqrt{3p\left(  1+12p\right)  }%
}~,~V\left(  t\right)  =\frac{6p\phi_{0}^{2}}{B_{0}}\left(  3\left(
1-8\right)  p-B_{0}-4\sqrt{3p\left(  1+12p\right)  }\right)  \phi^{-2\left(
1+12p\right)  -4\sqrt{3p\left(  1+12p\right)  }},
\end{equation}
that is%
\begin{equation}
V\left(  \phi\left(  t\right)  \right)  =\frac{6p\phi_{0}^{2-\sqrt
{\frac{1+12p}{3p}}}}{B_{0}}\left(  3\left(  1-8\right)  p-B_{0}-4\sqrt
{3p\left(  1+12p\right)  }\right)  \phi^{\sqrt{\frac{1+12p}{3p}}}.
\end{equation}
We take linear perturbations around the exact solution as before and for the perturbations we find
$\delta a\simeq t^{R}~,~\delta\phi=t^{S}$, where%
\begin{align}
R  &  =1+2\left(  12-\frac{1}{B_{0}}\right)  p+4\sqrt{3p\left(  1+12p\right)
},\\
S  &  =1+3\left(  4-\frac{1}{B_{0}}\right)  p+2\sqrt{3p\left(  1+12p\right)
},
\end{align}
from which we infer that the scaling solution is attractor when $0<B_{0}%
<\frac{1}{24}$, for $p>\frac{B_{0}+2\sqrt{6}B_{0}^{3/2}}{\left(
1-24B_{0}\right)  }$.

We proceed with the presentation of the analytic solutions.

\section{Analytic Solutions}

\label{sec4}

For the lapse function $N=1$, and when dust fluid is included in the model,
the point-like Lagrangian (\ref{ac.05}) is written as%
\begin{equation}
L\left(  a,\dot{a},\phi,\dot{\phi}\right)  =-3B\left(  \phi\right)  a\dot
{a}^{2}+\frac{1}{2}a^{3}\dot{\phi}^{2}-a^{3}V\left(  \phi\right)  -\rho_{m0},
\label{ac.19}%
\end{equation}
which describes the motion of a particle in a two-dimensional space, where now
the constraint equation (\ref{ac.09}) correspond to the Hamiltonian
conservation law for Lagrangian (\ref{ac.19}) with value the $\rho_{m0}$. The
equation of motions depend on two unknown functions, the $B\left(
\phi\right)  $ and the $V\left(  \phi\right)  $. Function $V\left(
\phi\right)  $ is a potential term, while function $B\left(  \phi\right)  $
defines the geometry of the two-dimensional space where the motion of the
point-like particle occurs.

The authors of \cite{Kanno:2006ty} considered the function $B\left(\phi\right)$ 
in the particular form $B\left(  \phi\right)  =6B_{0}\phi^{2}$, and, in our work,
this specific function will be selected as well. The reason is that
$B\left(\phi\right)=B_{0}\phi^{2}$ simplifies the dynamics such that the
minisuperspace defined by the kinetic part of Lagrangian (\ref{ac.19}), will be
a maximally symmetric two-dimensional space with zero curvature, that is, a
two-dimensional flat space. Therefore, the field equations describes a typical
dynamical system of Classical Mechanics.

As we commented before, we follow~\cite{Kanno:2006ty} and we set $V\left(  \phi\right)  =V_{0}\phi
^{2}$. For that specific functional forms of $B\left(  \phi\right)  $ and
$V\left(  \phi\right)  $, the field equations are written
\begin{equation}
18B_{0}\phi^{2}H^{2}-\frac{1}{2}\dot{\phi}^{2}-V_{0}\phi^{2}-\rho_{m0}%
a^{-3}=0, \label{ac.20}%
\end{equation}%
\begin{equation}
6B_{0}\phi^{2}\left(  2\dot{H}+3H^{2}\right)  +\frac{1}{2}\dot{\phi}%
^{2}+24\phi\dot{\phi}H-V_{0}\phi^{2}=0, \label{ac.21}%
\end{equation}%
\begin{equation}
\ddot{\phi}+3\dot{\phi}H+36B_{0}\phi H^{2}+2V_{0}\phi=0. \label{ac.22}%
\end{equation}

We define the canonical variables%
\begin{equation}
a=x^{\frac{1}{3-12\sqrt{B_{0}}}}y^{\frac{1}{3+12\sqrt{B_{0}}}}~,~\phi
=x^{\frac{2\left(  4B_{0}+\sqrt{B_{0}}\right)  }{16B_{0}-1}}y^{\frac{2\left(
4B_{0}-\sqrt{B_{0}}\right)  }{16B_{0}-1}}~,~B_{0}\neq\frac{1}{16}
\label{ac.23}%
\end{equation}
such that equations (\ref{ac.20}), (\ref{ac.21}), (\ref{ac.22}) are written in
simpler expressions as follows%
\begin{equation}
\frac{8B_{0}}{16B_{0}-1}\dot{x}\dot{y}+V_{0}xy+\rho_{m0}=0, \label{ac.24}%
\end{equation}%
\begin{equation}
\ddot{x}+2V_{0}\left(  1-\frac{1}{16B_{0}}\right)  x=0, \label{ac.25}%
\end{equation}%
\begin{equation}
\ddot{y}+2V_{0}\left(  1-\frac{1}{16B_{0}}\right)  y=0, \label{ac.26}%
\end{equation}
from which we derive the analytic solution in closed-form functions%
\begin{equation}
x\left(  t\right)  =x_{0}\sinh\left(  \omega t+x_{1}\right)  ,~y\left(
t\right)  =y_{0}\sinh\left(  \omega t+y_{2}\right)  , \label{ac.27}%
\end{equation}
with constraint $\rho_{m0}=-V_{0}x_{0}y_{0}\cosh\left(  x_{1}-y_{1}\right)  $,
and $\omega^{2}=V_{0}\left(  \frac{1}{8B_{0}}-2\right)  $. Because $V_{0}$ is
positive, we conclude that when $B_{0}>\frac{1}{16}$ we have a bounced
universe, while when $0<B<\frac{1}{16}$ the scale factor $a\left(  t\right)  $
is described by hyperbolic functions.

Consider now the special case where $x_{1}=y_{1}=0,$ then we find the scale
factor%
\begin{equation}
a\left(  t\right)  =a_{0}\left(  \sinh\left(  \omega t\right)  \right)
^{\frac{2}{3-48B_{0}}}.
\end{equation}
From the latter scale factor we see that the $\Lambda$-cosmology is not
recovered since $\frac{2}{3-48B_{0}}=\frac{2}{3}$ gives that $B_{0}=0$ which
is neglected. However when $B=\frac{3\varepsilon}{16\left(  3\varepsilon
+2\right)  }$ it follows $\frac{2}{3-48B_{0}}=\frac{2}{3}+\varepsilon$, hence
for small values of $\varepsilon$ or $B\simeq\frac{3}{32}\varepsilon$, the
exact solution is $a\left(  t\right)  =a_{0}\left(  \sinh\left(  \omega
t\right)  \right)  ^{\frac{2}{3}+\varepsilon},~$from which we calculate the
Hubble function%
\begin{equation}
H\left(  a\right)  =\frac{\omega}{3}\left(  2+3\varepsilon\right)
\sqrt{1+\left(  \frac{a_{0}}{a}\right)  ^{\frac{6}{2+3\varepsilon}}},
\end{equation}
which is the solution of General Relativity of an ideal gas with a
cosmological constant term. Recall, that the field $\phi$ contributes in the
cosmological fluid while function $B\left(\phi\right)$ affect the total
fluid source.

When $B_{0}=\frac{1}{16}$, we introduce the new canonical variables%
\begin{equation}
a=u^{\frac{1}{6}}e^{\frac{v}{2}}~,~\phi=u^{\frac{1}{4}}e^{-\frac{3}{4}v},
\label{ac.28}%
\end{equation}
where the field equations take the form%
\begin{equation}
\frac{3}{8}\dot{u}\dot{v}-V_{0}u-\rho_{m0}=0 \label{ac.29}%
\end{equation}%
\begin{equation}
\ddot{u}=0,~\ddot{v}=\frac{8}{3}V_{0}, \label{ac.30}%
\end{equation}
with solution $u\left(  t\right)  =u_{1}t+u_{0}~,~v\left(  t\right)  =\frac
{4}{3}V_{0}t^{2}+v_{1}t+v_{0}$ with constraint equation $\rho_{m0}+V_{0}%
u_{1}-3u_{1}v_{1}=0$, from which we find the scale factor $a\left(  t\right)
=a_{0}t^{\frac{1}{6}}e^{\frac{2V_{0}}{3}t^{2}+\frac{v_{1}}{2}}$.

When the scalar field is massless, i.e. $V\left(  \phi\right)  =0$, then field
equations are reduced, and the generic solution can be easily
constructed by equations (\ref{ac.25}), (\ref{ac.26}) for $V_{0}=0$ and the
transformation rule (\ref{ac.23}).

These are not the only functional forms of the potential $V\left(  \phi\right)$
for which we can write the analytic solutions of the field equations. Some
power-law functions $V\left(\phi\right)$ and their analytic solutions are
presented in what it follows. Specifically, the potentials for which we shall
present the analytic solution of the field equations are%
\begin{equation}
V_{A}\left(  \phi\right)  =V_{0}\phi^{2}+V_{1}\phi^{\frac{1}{2\sqrt{B_{0}}}}~,
\label{ac.31}%
\end{equation}%
\begin{equation}
V_{B}\left(  \phi\right)  =V_{0}\phi^{2}+V_{1}\phi^{-\frac{1}{2\sqrt{B_{0}}}},
\label{ac.32}%
\end{equation}%
\begin{equation}
V_{C}\left(  \phi\right)  =V_{0}\phi^{2}+V_{1}\phi^{-2+\frac{1}{4B_{0}}},
\label{ac.33}%
\end{equation}%
\begin{equation}
V_{D}\left(  \phi\right)  =V_{0}\phi^{\frac{1}{2\sqrt{B_{0}}}}+V_{1}%
\phi^{-1+\frac{3}{4\sqrt{B_{0}}}}, \label{ac.34}%
\end{equation}%
\begin{equation}
V_{E}\left(  \phi\right)  =V_{0}\phi^{-\frac{1}{2\sqrt{B_{0}}}}+V_{1}%
\phi^{-1-\frac{3}{4\sqrt{B_{0}}}}. \label{ac.35}%
\end{equation}

Potentials (\ref{ac.31})-(\ref{ac.35}) have a common feature, they are 
Liouville--integrable, for which the field equations admit an additional
conservation law for each potential, more specifically a quadratic
conservation law, different for each potential. The $V\left(  \phi\right)
=V_{0}\phi^{2}$ is also a super-integrable potential. Another superintegrable
potentials we observe are $\bar{V}_{A}\left(  \phi\right)  =V_{1}%
\phi^{\frac{1}{2\sqrt{B_{0}}}}$ and $\bar{V}_{B}\left(  \phi\right)
=V_{1}\phi^{-\frac{1}{2\sqrt{B_{0}}}}$. Integrable cosmological models in modified 
theories of gravity have been widely studied in the literature, and have been drawn 
the attention of cosmologists and of the mathematicians ever. The main reason is 
that analytic solutions can be used as toy models in order to understand the main 
properties of a proposed cosmological theory
\cite{ns01,ns02,ns03,ns04,ns05,ns06,ns07,ns08,ns09,ns10,ns11}.

In order to describe Nature, we need a large number of free
parameters or boundary conditions, which makes numerical treatment worthless. 
Because in general the equations which describe a specific theory are nonlinear 
such that numerical solutions may be sensitive on small changes of the initial 
conditions. Consequently, we refer to analytic techniques in order to understand 
the generic properties of a propose physical theory. Hence, the knowledge for the 
existence and the determination of analytical or exact solutions for a given 
dynamical system is important for the detailed study and understanding of the given physical theory.

According to the results of the previous Section, we observe that according to
which term of the potential dominates, the behavior of the scale factor will
be near to that of the scaling solution or to that of an exponential solution. For instance
consider the potential $V_{B}\left(\phi\right)$. For large values of
$\phi$, it follows $V_{B}\left(\phi\right)  \simeq\phi^{2}$ from which we
infer that solution approaches the de Sitter universe, on the other hand, for
small values of $\phi$, $V_{B}\left(  \phi\right)  \simeq\phi^{-\frac
{1}{2\sqrt{B_{0}}}}$, from which we infer that the scale factor behaves like
that of the scaling solution. As far as potentials $V_{D}\left(\phi\right)
,~V_{E}\left(\phi\right)  $, are concerned, we remark that they have two
terms, where only scaling solutions are described.

The method that we apply in order to determine the analytic solutions is based
on canonical coordinates, as described in the example $V\left(  \phi\right)
=V_{0}\phi^{2}$. \ In the following, we present the analytic solutions.

\subsection{Analytic solution for potential $V_{A}\left(  \phi\right)  $}

For the potential $V_{A}\left(  \phi\right)  ,$ and for $B_{0}\neq\frac{1}%
{16}$ we find the generic solution in the canonical coordinates $\left\{
x,y\right\}  $ defined by expression (\ref{ac.23})%
\begin{align}
y\left(  t\right)   &  =y_{1}e^{\omega t}+y_{2}e^{-\omega t},\label{ac.36}\\
x\left(  t\right)   &  =x_{1}e^{\omega t}+x_{2}e^{\omega t}+x_{sp}\left(
t\right),  \label{ac.37}%
\end{align}
where%
\begin{equation}
x_{sp}=-\frac{y_{2}V_{1}}{8\sqrt{B_{0}}V_{0}}\left(  y_{2}e^{\omega t}\right)
^{\left(  1-\frac{2}{1+4\sqrt{B_{0}}}\right)  }\left(  _{2}F_{1}\left(
\alpha,\beta,\gamma,\zeta\left(  t\right)  \right)  +4\sqrt{B_{0}}~_{2}%
F_{1}\left(  \alpha^{\prime},\beta^{\prime},\gamma^{\prime},\zeta\left(
t\right)  \right)  \right),  \label{ac.38}%
\end{equation}
where $_{2}F_{1}\left(  \alpha,\beta,\gamma,\zeta\left(  t\right)  \right)  $
denote the hypergeometric function represented by the hypergeometric series;
while $\alpha=1-\frac{1}{1+4\sqrt{B_{0}}},~\beta=1-\frac{2}{1+4\sqrt{B_{0}}%
},~\gamma=1+\alpha,~\zeta\left(  t\right)  =-\frac{y_{1}}{y_{2}}e^{\omega t}$
and $\alpha^{\prime}=\beta,~\beta^{\prime}=a-1,~\gamma^{\prime}=\alpha$, where
$\omega^{2}=V_{0}\left(  \frac{1}{8B_{0}}-2\right)  $ and constraint~$\rho
_{m0}+2\left(  y_{1}x_{2}+y_{2}x_{1}\right)  =0$.

In the special case where $y_{2}=0$ \ exact solution is simplified
\begin{align}
y\left(  t\right)   &  =y_{1}e^{\omega t},\label{ac.39}\\
x\left(  t\right)   &  =x_{1}e^{\omega t}+x_{2}e^{-\omega t}-\frac
{1+4\sqrt{B_{0}}}{8\sqrt{B_{0}}V_{0}}V_{1}\left(  y_{1}e^{\omega t}\right)
^{\left(  1-\frac{2}{1+4\sqrt{B_{0}}}\right)  }. \label{ac.40}%
\end{align}
For the latter two solutions, namely (\ref{ac.36}), (\ref{ac.37}) and
(\ref{ac.39}), (\ref{ac.40}) when $B_{0}<\frac{1}{16},~$that is, $1-\frac
{2}{1+4\sqrt{B_{0}}}<0$, for large time the dominated term is $e^{\omega t}$,
which means that the scale factor for large values of $t$,
approaches that of the de Sitter universe $a\left(t\right)=a_{0}e^{H_{0}t}$.

\begin{figure}[ptb]
\centering\includegraphics[width=0.8\textwidth]{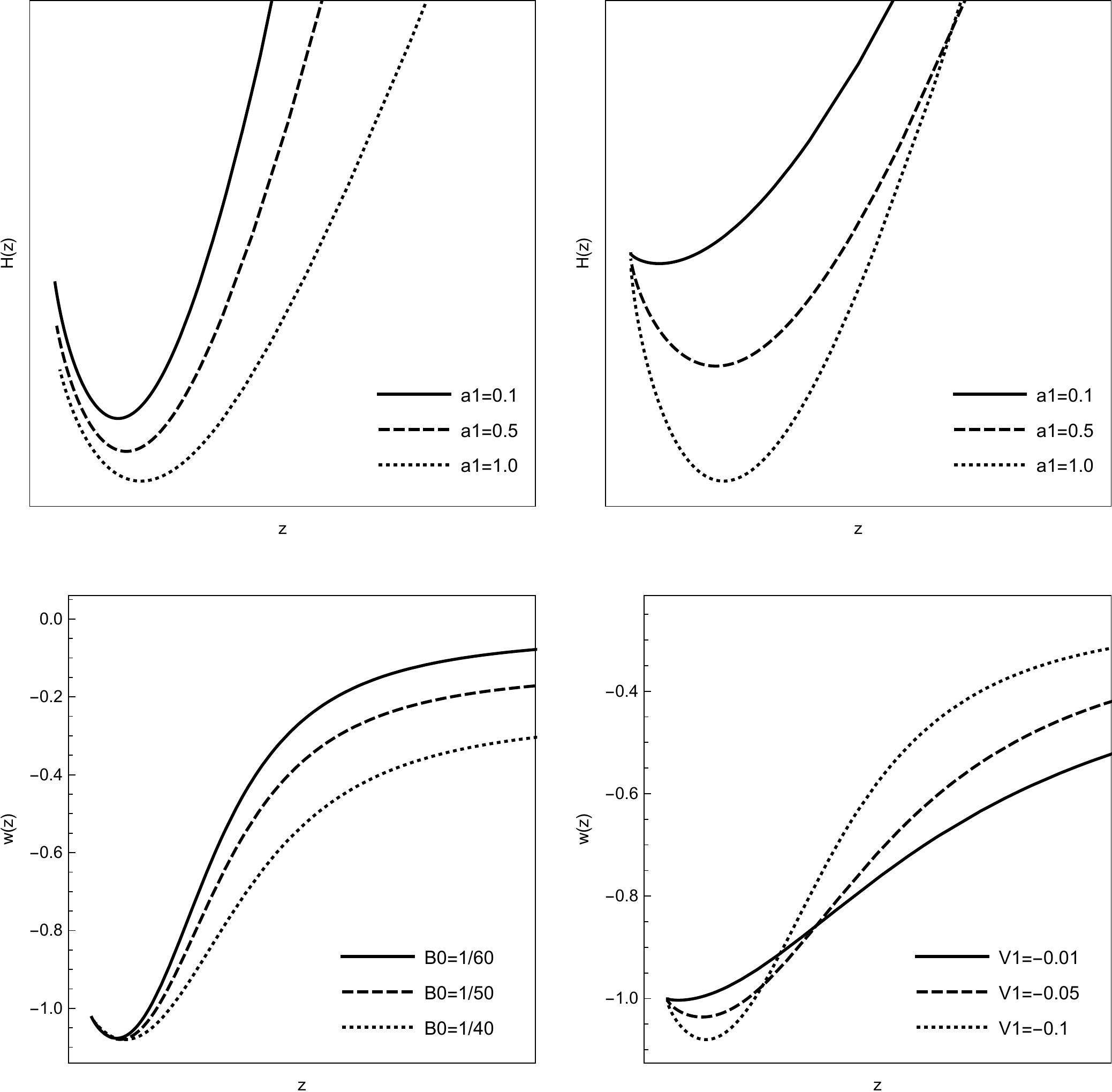} \newline%
\caption{Qualitative evolution for the Hubble function $H\left(  z\right)  $
and for the equation of state parameter $w\left(  z\right)  $ for the
total fluid for the analytic solution (\ref{ac.39}), (\ref{ac.40}). For the
figures we considered, $\left(  x_{1},x_{2},y_{1},V_{0}\right)  =\left(
0.1,-0.1,1,0.5\right)  $. Left figures are for $V_{1}=-0.1$ and different
values of $B_{0}=\left\{  \frac{1}{60},\frac{1}{50},\frac{1}{40}\right\}  $.
Right figures are for $B_{0}=\frac{1}{40}$ and different values of
$V_{1}=\left\{  -0.01,-0.05,-0.1\right\}  $. From the plots we observe that
the cosmological fluid pass the phantom divide line while the de Sitter
universe is a future attractor. }%
\label{fig1}%
\end{figure}

For the latter exact solution (\ref{ac.39}), (\ref{ac.40}), in Fig. \ref{fig1}
we present the qualitative behaviour of the Hubble function $H\left(
z\right)  $ and of the parameter for the equation of state for the effective
fluid $w\left(  z\right)  $ in terms of the redshift $1+z=\frac{1}{a}$.

Now, if we assume that $V_{0}=0$, then for the super-integrable potential
$\bar{V}_{A}\left(  \phi\right)  $ we find the exact solution%
\begin{align}
y\left(  t\right)   &  =y_{1}t+y_{2},\label{ac.41}\\
x\left(  t\right)   &  =x_{1}t+x_{2}+\frac{\left(  1+4\sqrt{B_{0}}\right)
^{2}}{8B_{0}y_{1}^{2}}V_{1}\left(  1-4\sqrt{B_{0}}\right)  \left(
y_{1}t+y_{2}\right)  ^{1+\frac{2}{1+4\sqrt{B_{0}}}}, \label{ac.42}%
\end{align}
with constraint equation $\rho_{m0}-\frac{8B_{0}}{1-16B_{0}}y_{1}x_{1}=0.$ For
solution (\ref{ac.41}), (\ref{ac.42}) for large values of time the scale
factor has a power-law behaviour $a\left(  t\right)  =a_{0}t^{p}$, where
$p=p\left(  B_{0}\right)  $.

When $B_{0}=\frac{1}{16}$, potential $V_{A}\left(  \phi\right)  $ reads
$V_{A}\left(  \phi\right)  =\left(  V_{0}+V_{1}\right)  \phi^{2}$, which was
the one studied before. Hence, we continue by presenting the analytic solution
for potential $V_{B}\left(  \phi\right)  $.

\subsection{Potential $V_{B}\left(  \phi\right)  $}

For the second potential of our consideration, namely potential $V_{B}\left(
\phi\right)  $, and for $B_{0}\neq\frac{1}{16}$, we find the generic analytic
solution written in closed-form expression%
\begin{align}
x\left(  t\right)   &  =x_{1}e^{\omega t}+x_{2}e^{-\omega t},\label{ac.43}\\
y\left(  t\right)   &  =y_{1}e^{\omega t}+y_{2}e^{-\omega t}+y_{sp}\left(
t\right), \label{ac.44}%
\end{align}
in which
\begin{equation}
y_{sp}\left(  t\right)  =\frac{\left(  4\sqrt{B_{0}}-1\right)  ^{2}}%
{64B_{0}^{3/2}\omega^{2}x_{2}}V_{1}e^{\omega t}\left(  \frac{\left(
1+\frac{x_{1}}{x_{2}}e^{2\omega t}\right)  }{x_{1}e^{\omega t}+x_{2}e^{-\omega
t}}\right)  ^{\frac{2}{4\sqrt{B_{0}}-1}}\left(  4\sqrt{B_{0}}_{2}F_{1}\left(
\bar{\alpha},\bar{\beta},\bar{\gamma},\bar{\zeta}\left(  t\right)  \right)
-_{2}F_{1}\left(  \bar{a}^{\prime},\bar{\beta}^{\prime},\bar{\gamma}^{\prime
},\bar{\zeta}\left(  t\right)  \right)  \right)  , \label{ac.45}%
\end{equation}
in which $\zeta\left(  t\right)  =-\frac{x_{1}}{x_{2}}e^{2\omega t}%
,~\bar{\alpha}=1-\frac{2}{1-4\sqrt{B_{0}}},~\bar{\beta}=\frac{\bar{a}+1}%
{2},~\bar{\gamma}=4\sqrt{B_{0}}\bar{\beta},~\bar{\alpha}^{\prime}=1+\bar
{\beta},~\bar{\beta}^{\prime}=\bar{a}$ and $\bar{\gamma}^{\prime}=1+\bar{a}$.
\ Furthermore, from the constraint equation it follows the algebraic condition
$\rho_{m0}+2\left(  y_{1}x_{2}+y_{2}x_{1}\right)  =0$.

If $x_{2}=0$, the closed-form solution is
\begin{align}
x\left(  t\right)   &  =x_{1}e^{\omega t},\label{ac.46}\\
y\left(  t\right)   &  =y_{1}e^{\omega t}+y_{2}e^{-\omega t}+\frac{\left(
4\sqrt{B_{0}}-1\right)  ^{3}}{64B_{0}^{3/2}\omega^{2}}V_{1}\left(
x_{1}e^{\omega t}\right)  ^{-1+\frac{2}{1-4\sqrt{B_{0}}}}, \label{ac.47}%
\end{align}
from which we infer, similarly as for the potential $V_{A}\left(  \phi\right)$, that
the scale factor for large values of $t$, it is approximated by that of the de
Sitter universe.

When $V_{0}=0$, the analytic solution is found to be%
\begin{align}
x\left(  t\right)   &  =x_{1}t+x_{2},\label{ac.48}\\
y\left(  t\right)   &  =y_{1}t+y_{2}+\frac{\left(  4\sqrt{B_{0}}-1\right)
^{3}V_{1}}{\left(  4\sqrt{B_{0}}-3\right)  8B_{0}x_{1}^{2}}\left(
x_{1}t+x_{2}\right)  ^{1+\frac{2}{1-4\sqrt{B_{0}}}}. \label{ac.49}%
\end{align}
with constraint condition~$\rho_{m0}-\frac{8B_{0}}{1-16B_{0}}y_{1}x_{1}=0$,
while when $x_{1}=0$ the generic analytic solution is
\begin{align}
x\left(  t\right)   &  =x_{2},\label{ac.50}\\
y\left(  t\right)   &  =y_{1}t+y_{2}+\frac{\left(  4\sqrt{B_{0}}-1\right)
}{8B_{0}}V_{1}\left(  x_{2}\right)  ^{-1+\frac{2}{1-4\sqrt{B_{0}}}}t^{2}.
\label{ac.51}%
\end{align}
The latter solutions are physically accepted if and only if $\rho_{m0}=0$, that
is, there is not any contribution by the dust fluid in the cosmological fluid.
Consequently from (\ref{ac.23}) we infer that the scale factor for the
solutions with $V_{0}=0$ have a power-law expression.

For $B=\frac{1}{16}$, we work with the variables $\left\{  u,v\right\}  $
which are defined by expression (\ref{ac.28}). Hence, the field equations are
reduced to the following system%
\begin{align}
\ddot{u}-8V_{1}e^{3v}  &  =0,\label{ac.52}\\
\ddot{v}-\frac{8}{3}V_{0}  &  =0, \label{ac.53}%
\end{align}
from where it follows the generic solutions%
\begin{align}
v\left(  t\right)   &  =\frac{4}{3}V_{0}t^{2}+v_{1}t+v_{2},\label{ac.54}\\
u\left(  t\right)   &  =u_{1}t+u_{2}+\frac{e^{3v\left(  t\right)  }%
V_{1}\left(  \left(  8tV_{0}+3V_{1}\right)  D_{+}\left(  \frac{8V_{0}t+3V_{1}%
}{4\sqrt{V_{0}}}\right)  -2\sqrt{V_{0}}\right)  }{2V_{0}}, \label{ac.55}%
\end{align}
where $D_{+}\left(  t\right)  $ is the Dawson function defined as
$D_{+}\left(  t\right)  =e^{-t^{2}}\int_{0}^{t}e^{r^{2}}dr$, while from the
constraint equation it follows $\rho_{m0}=\frac{3}{8}v_{1}u_{1}$. Last, but not
least, when $V_{0}=0$, i.e. $V_{B}\left(  \phi\right)  =V_{1}\phi^{-\frac
{1}{2\sqrt{B_{0}}}}$ the analytic solution is
\begin{align}
v\left(  t\right)   &  =v_{1}t+v_{2},\label{ac.56}\\
u\left(  t\right)   &  =u_{1}t+u_{2}+\frac{8V_{1}}{9v_{1}^{2}}e^{3\left(
v_{1}t+v_{2}\right)  }. \label{ac.57}%
\end{align}
From solution (\ref{ac.56}), (\ref{ac.57}) we calculate
\begin{equation}
a\left(  t\right)  =a_{0}\left(  e^{\frac{12}{5}v_{1}t}+a_{1}te^{-\frac{3}%
{5}v_{1}t}\right)  ^{\frac{5}{12}}, \label{ac.58}%
\end{equation}
which gives%
\begin{equation}
H\left(  t\right)  =v_{1}+\frac{5a_{1}\left(  1-3tv_{1}\right)  }%
{12e^{3v_{1}t}+a_{1}t}, \label{ac.59}%
\end{equation}
which means that for large values of $t$ and for positive $v_{1}$, the
solution behaves like that of the de\ Sitter universe.

\begin{figure}[ptb]
\centering\includegraphics[width=0.8\textwidth]{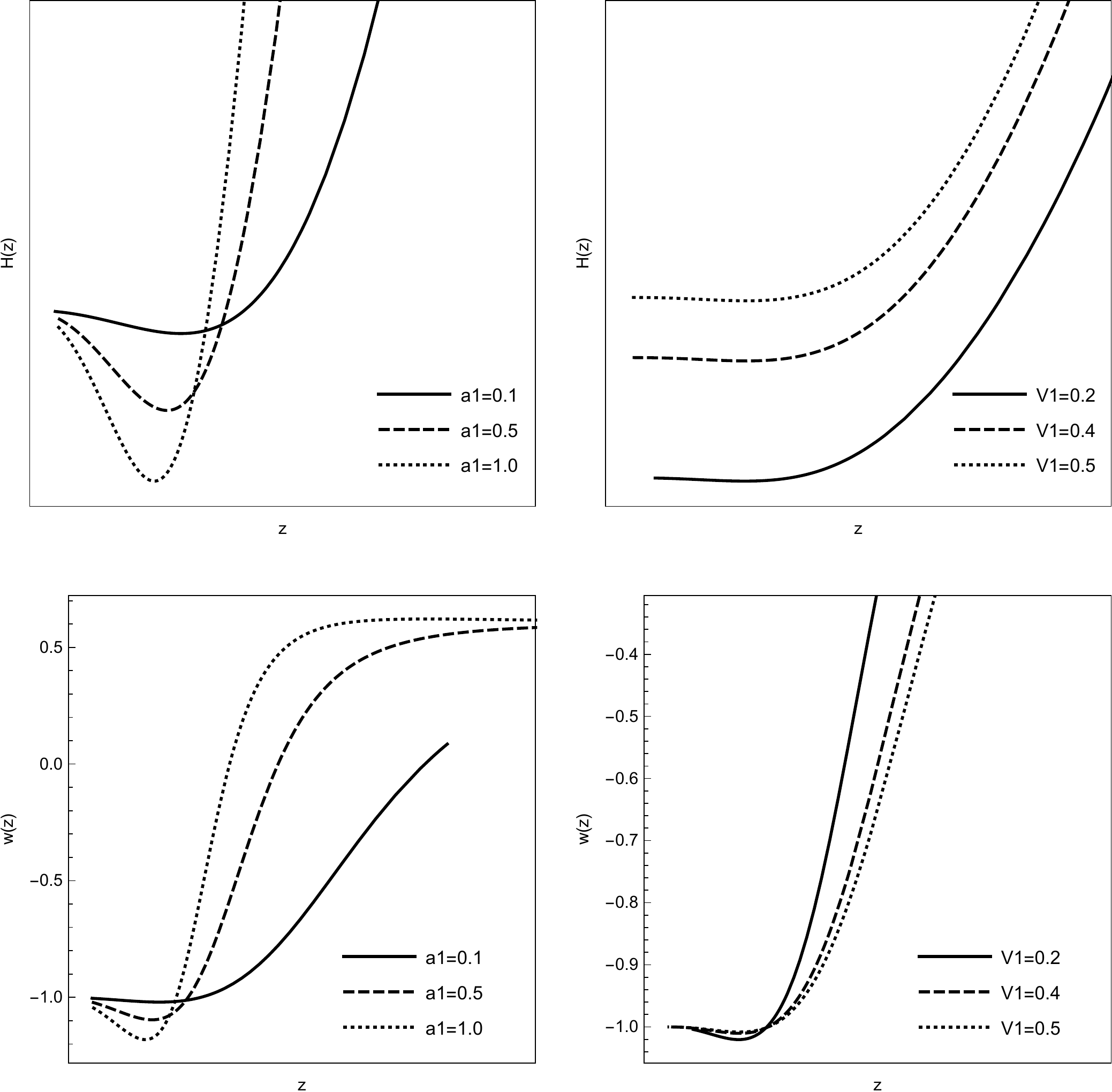} \newline%
\caption{Qualitative evolution for the Hubble function $H\left(  z\right)  $
and for the equation of state parameter $w\left(  z\right)  $ for the
total fluid for the scale factor (\ref{ac.59}) which holds for $B_{0}=\frac
{1}{16}$. For the figures we considered \thinspace$a_{0}=1$. Left figures are
for $V_{1}=0.2$ and different values of $a_{1}=\left\{  -0.1,0.5,1\right\}  $.
Right figures are for $a_{1}=0.1$ and different values of $V_{1}=\left\{
0.2,0.4,0.5\right\}  $. }%
\label{fig2}%
\end{figure}

For the scale factor (\ref{ac.58}) we present in Fig. \ref{fig2}  the qualitative behavior of the
Hubble function $H\left(z\right)$, as well as the behavior of the equation of
state parameter $w\left(z\right)$ of the effective fluid in terms of the redshift
$1+z=\frac{1}{a}$.

\subsection{Potential $V_{C}\left(  \phi\right)  $}

For the potential $V_{C}\left(  \phi\right)$ we choose the
coordinates
\begin{align}
a  &  =r^{\frac{2}{3\left(  1-16B_{0}\right)  }}\left(  \cosh\theta
+\sinh\theta\right)  ^{\frac{1}{3-12\sqrt{B_{0}}}}\left(  \cosh\theta
-\sinh\theta\right)  ^{\frac{1}{3+12\sqrt{B_{0}}}}~,\label{ac.60}\\
~\phi &  =r^{1+\frac{1}{16B_{0}-1}}\left(  \cosh\theta+\sinh\theta\right)
^{\frac{2\left(  4B_{0}+\sqrt{B_{0}}\right)  }{16B_{0}-1}}\left(  \cosh
\theta-\sinh\theta\right)  ^{\frac{2\left(  4B_{0}-\sqrt{B_{0}}\right)
}{16B_{0}-1}}~,~B_{0}\neq\frac{1}{16}. \label{ac.61}%
\end{align}
In the new coordinates the point-like Lagrangian reads
\begin{equation}
L_{C}\left(  r,\dot{r},\theta,\dot{\theta}\right)  =\frac{8B_{0}}{1-16B_{0}%
}\left(  -\dot{r}^{2}+r^{2}\dot{\theta}^{2}\right)  -V_{0}r^{2}-V_{1}%
\frac{e^{-\frac{\theta}{\sqrt{B_{0}}}}}{r^{2}}-\rho_{m0}. \label{ac.62}%
\end{equation}
Easily we observe that it is the Ermakov-Pinney system defined in the
two-dimensional space of Lorentzian signature.

The Hamiltonian function is written as follows%
\begin{equation}
\frac{1}{2}\left(  -\dot{r}^{2}+r^{2}\dot{\theta}^{2}\right)  -\bar{V}%
_{0}r^{2}+\bar{V}_{1}\frac{e^{-\frac{\theta}{\sqrt{B_{0}}}}}{r^{2}}+\bar{\rho
}_{m0}=0, \label{ac.63}%
\end{equation}
where $\left(  V_{0},V_{1},\rho_{m0}\right)  =\frac{4B_{0}}{1-16B_{0}}\left(
-\bar{V}_{0},\bar{V}_{1},\bar{\rho}_{m0}\right)  .$ By using the momentum,
$p_{r}=\frac{\partial L}{\partial\dot{r}}$ and $p_{\theta}=\frac{\partial
L}{\partial\theta}$, the constraint equation becomes%
\begin{equation}
-\frac{1}{2}p_{r}^{2}+\bar{V}_{0}r^{2}+\bar{\rho}_{m0}+\frac{p_{\theta}%
^{2}+2\bar{V}_{1}e^{-\frac{\theta}{\sqrt{B_{0}}}}}{2r^{2}}=0, \label{ac.64}%
\end{equation}
from which we infer the two first-order ordinary differential equations%
\begin{align}
\frac{1}{2}p_{\theta}^{2}+\bar{V}_{1}e^{-\frac{\theta}{\sqrt{B_{0}}}}  &
=J_{0},\label{ac.65}\\
-\frac{1}{2}p_{r}^{2}+\bar{V}_{0}r^{2}+\bar{\rho}_{m0}+\frac{J_{0}}{r^{2}}  &
=0. \label{ac.66}%
\end{align}

Thus the generic solution is
\begin{equation}
r^{2}\left(  t\right)  =r_{1}e^{2\sqrt{2V_{0}}t}+r_{2}e^{-2\sqrt{2V_{0}}%
t}+r_{3}, \label{ac.67}%
\end{equation}
with constraints $V_{0}r_{3}^{2}-\left(  4r_{1}r_{2}V_{0}-J_{0}V_{1}\right)  $
and $\rho_{m0}=2\left\vert V_{0}r_{3}\right\vert $, while $\theta\left(
t\right)  $ is given by the first-order ordinary differential equation%
\begin{equation}
\frac{1}{2}r^{2}\dot{\theta}^{2}+V_{1}e^{-\frac{\theta}{\sqrt{B_{0}}}}%
-J_{0}=0. \label{ac.68}%
\end{equation}

When $V_{0}=0$, the exact solution is
\begin{equation}
r\left(  t\right)  =r_{1}\left(  t-t_{0}\right)  ^{2}+r_{3}\left(
t-t_{0}\right)  \label{ac.69}%
\end{equation}
with constraints $8V_{1}J_{0}=-r_{3}^{2}$ and $2\rho_{m1}+r_{1}=0$, while
$\theta\left(  t\right)  $ is given again by equation (\ref{ac.68}). Remark
that for $B_{0}=\frac{1}{16}$ it follows $V_{C}\left(  \phi\right)  =\left(
V_{0}+V_{1}\right)  \phi^{2}$.

\subsection{Potential $V_{D}\left(  \phi\right)  $}

For potential $V_{D}\left(  \phi\right)  $, in the canonical coordinates
$\left\{  x,y\right\}  $ the constraint equation, i.e. the Hamiltonian of the
dynamical system is written as
\begin{equation}
p_{x}p_{y}-\bar{V}_{0}y^{\frac{2}{1+4\sqrt{B_{0}}}}-\bar{V}_{1}\frac
{y^{-\frac{1}{2}+\frac{3}{1+4\sqrt{B_{0}}}}}{\sqrt{x}}-\bar{\rho}_{m0}=0,
\label{ac.70}%
\end{equation}
where $\left\{  p_{x},p_{y}\right\}  =\left\{  \dot{y},\dot{x}\right\}  $ and
$\left(  \bar{V}_{0},\bar{V}_{1},\bar{\rho}_{m0}\right)  =\left(  \frac
{1}{8B_{0}}-2\right)  \left(  V_{0},V_{1},\rho_{m0}\right)  $. The dynamical
system admits the additional conservation law%
\begin{equation}
xp_{x}^{2}-yp_{x}p_{y}-\bar{V}_{1}\frac{y^{-\frac{1}{2}+\frac{3}%
{1+4\sqrt{B_{0}}}}}{\sqrt{x}}+\frac{2\bar{V}_{0}}{3+4\sqrt{B_{0}}}%
y^{1+\frac{2}{1+4\sqrt{B_{0}}}}=I_{0}. \label{ac.71}%
\end{equation}

The Action which follows as a solution of the Hamilton Jacobi equation is
calculated as
\begin{equation}
S_{D}\left(  x,y\right)  =2\sqrt{x\left(  \bar{\rho}_{m0}y+I_{0}\right)
+\frac{1+4\sqrt{B_{0}}}{3+4\sqrt{B_{0}}}\bar{V}_{0}xy^{1+\frac{2}%
{1+4\sqrt{B_{0}}}}}+\bar{V}_{1}\bigint\frac{y^{-\frac{-5+4\sqrt{B_{0}}}{2\left(
1+4\sqrt{B_{0}}\right)  }}}{\sqrt{\left(  \rho_{m0}y+I_{0}\right)
+\frac{1+4\sqrt{B_{0}}}{3+4\sqrt{B_{0}}}\bar{V}_{0}y^{1+\frac{2}%
{1+4\sqrt{B_{0}}}}}}dy, \label{ac.72}%
\end{equation}
such that the analytic solution of the field equations is given by the
following system of two first-order ordinary differential equations%
\begin{equation}
\dot{x}=\frac{\partial S_{D}}{\partial y}~,~\dot{y}=\frac{\partial S_{D}%
}{\partial x}, \label{ac.73}%
\end{equation}
where in the special case where $\bar{\rho}_{m0}=0,~\bar{V}_{0}=0$ it becomes%
\begin{equation}
\dot{x}=\frac{\bar{V}_{1}}{\sqrt{I_{0}}}y^{-\frac{-5+4\sqrt{B_{0}}}{2\left(
1+4\sqrt{B_{0}}\right)  }}~,~\dot{y}=\sqrt{\frac{I_{0}}{x}}, \label{ac.74}%
\end{equation}
or equivalently,%
\begin{equation}
x^{\frac{3}{2}}=\frac{2I_{0}\bar{V}_{1}\left(  1+4\sqrt{B_{0}}\right)
}{7+4\sqrt{B_{0}}}y^{1-\frac{-5+4\sqrt{B_{0}}}{2\left(  1+4\sqrt{B_{0}%
}\right)  }}+c_{0},
\end{equation}
or%
\begin{equation}
I_{0}^{3/2}\ddot{y}+\bar{V}_{1}y^{-\frac{-5+4\sqrt{B_{0}}}{2\left(
1+4\sqrt{B_{0}}\right)  }}\dot{y}^{3}=0. \label{ac.75}%
\end{equation}
A special solution of the latter equation is the power-law expression $y\simeq
t^{p},~p=\frac{2}{3}\frac{\left(  1+3\sqrt{B_{0}}\right)  }{3+3\sqrt{B_{0}}}$,
which leads to a power law scale factor. We remark that when $B_{0}=\frac{1}{16}$ for the potential $V_{D}\left(
\phi\right)  ~$it follows $V_{D}\left(  \phi\right)  =\left(  V_{0}%
+V_{1}\right)  \phi^{2}$.

\subsection{Potential $V_{E}\left(  \phi\right)  $}

The procedure that we follow to write the analytic solution for potential
$V_{E}\left(  \phi\right)  $ is based on the derivation of the Action by
solving the Hamilton-Jacobi equation, as we did for potential $V_{D}\left(
\phi\right)  $.

In the canonical coordinates $\left\{  x,y\right\}  $ the constraint equation
reads%
\begin{equation}
p_{x}p_{y}-\bar{V}_{1}\frac{x^{-\frac{1}{2}+\frac{3}{1-4\sqrt{B_{0}}}}}%
{\sqrt{y}}-\bar{V}_{0}x^{\frac{2}{1-4\sqrt{B_{0}}}}-\bar{\rho}_{m0}=0,
\label{ac.76}%
\end{equation}
where $\left(  \bar{V}_{0},\bar{V}_{1},\bar{\rho}_{m0}\right)  =\left(
\frac{1}{8B_{0}}-2\right)  \left(  V_{0},V_{1},\rho_{m0}\right)  $. The
quadratic conservation law admitted by the field equations is%
\begin{equation}
yp_{y}^{2}-xp_{x}p_{y}+\bar{V}_{1}\frac{x^{\frac{1}{2}+\frac{3}{1-4\sqrt
{B_{0}}}}}{\sqrt{y}}+\frac{2\bar{V}_{0}}{3-4\sqrt{B_{0}}}x^{1+\frac
{2}{1-4\sqrt{B_{0}}}}=I_{0}. \label{ac.77}%
\end{equation}

Consequently, the Action is calculated as%
\begin{equation}
S_{E}\left(  x,y\right)  =\sqrt{y\left(  \bar{\rho}_{m0}x+I_{0}\right)
+\frac{1-4\sqrt{B_{0}}}{3-4\sqrt{B_{0}}}x^{\frac{3-4\sqrt{B_{0}}}%
{1-4\sqrt{B_{0}}}}}+\bar{V}_{1}\bigint\frac{x^{-\frac{5+4\sqrt{B_{0}}}{2\left(
1-4\sqrt{B_{0}}\right)  }}}{\sqrt{\left(  \bar{\rho}_{m0}x+I_{0}\right)
+\frac{1-4\sqrt{B_{0}}}{3-4\sqrt{B_{0}}}x^{\frac{3-4\sqrt{B_{0}}}%
{1-4\sqrt{B_{0}}}}}}dx, \label{ac.78}%
\end{equation}
where the reduced equations are
\begin{equation}
\dot{x}=\frac{\partial S_{E}}{\partial y}~,~\dot{y}=\frac{\partial S_{E}%
}{\partial x}. \label{ac.79}%
\end{equation}
There are similarities of the latter solution with that of potential
$V_{D}\left(  \phi\right)  $, but for different value of the constant $B_{0}$,
specifically by replacing mathematically $B_{0}\rightarrow i^{4}B_{0}$.

For $B_{0}=\frac{1}{16}$ in the canonical coordinates $\left\{  u,v\right\}  $
the constraint equation becomes%
\begin{equation}
\frac{3}{8}p_{u}p_{v}-V_{0}e^{3v}-V_{1}\frac{e^{\frac{9}{2}v}}{\sqrt{u}}%
-\rho_{m0}=0, \label{ac.80}%
\end{equation}
while the quadratic conservation law reads%
\begin{equation}
up_{u}^{2}-vp_{u}p_{v}+\frac{8}{3}V_{1}\frac{ve^{\frac{9}{2}v}}{\sqrt{u}%
}+\frac{8}{3}e^{3v}\left(  v-\frac{1}{3}\right)  -I_{0}=0, \label{ac.81}%
\end{equation}
in which $\left\{  \dot{u},\dot{v}\right\}  =\left\{  p_{v},p_{u}\right\}  $.
Hence, the Action is derived%
\begin{equation}
S_{E}\left(  u,v\right)  =-\frac{2}{3}\sqrt{u\left(  24u\rho_{m0}%
+9I_{0}+8V_{0}e^{3v}\right)  }-8V_{1}\bigint\frac{e^{\frac{9}{2}v}}{\sqrt{\left(
24u\rho_{m0}+9I_{0}+8V_{0}e^{3v}\right)  }}dv. \label{ac.82}%
\end{equation}

In the special case where $V_{0}=0$ and $\rho_{m0}$ the field equations reduce
to the simple form%
\begin{equation}
\dot{u}=\frac{8}{3}\frac{V_{1}}{\sqrt{I_{0}}}e^{\frac{9}{2}v}~,~\dot{v}%
=\sqrt{\frac{I_{0}}{u}}, \label{ac.83}%
\end{equation}
that is $u^{\frac{3}{2}}=\frac{1}{12}\frac{V_{1}}{I_{0}}e^{\frac{9}{2}y}$ or
equivalently%
\begin{equation}
\ddot{v}+\frac{4}{3}\frac{V_{1}}{\left(  I_{0}\right)  ^{\frac{3}{2}}}%
e^{\frac{9}{2}y}\dot{v}^{3}=0, \label{ac.84}%
\end{equation}
which is an integrable differential equation and admits the special solution
$e^{v}\simeq t^{\frac{2}{9}}.$ Finally, from (\ref{ac.28}) and (\ref{ac.83})
the scale factor is found to be a power-law function.

\section{Conclusions}

\label{sec5}

In this work we have considered a Lorentz--violating scalar field cosmological model
in a spatially flat FLRW background space. Specifically, we have included an aether field in the Einstein-Hilbert action leading to the
Einstein-aether theory, where the aether field is coupled to the scalar field through the aether parameters as proposed by Kanno et al. \cite{Kanno:2006ty}. 
The resulting field equations are of second-order, as expected for such kind of theories, and they can be produced by a point-like Lagrangian. There are similarities with 
scalar-tensor theories although they are quite different theories.

We have focused on the construction of scalar field potentials to see whether the field
equations are Liouville--integrable, that means that the field equations can be
solved in quadratures. Consequently, we investigated the functional forms of
the scalar field potential where the field equations admit conservation laws quadratic in the
momenta. By using the second conservation law we were able to write the analytic solution of the field equations for that specific scalar
field potentials and whenever it was feasible, we have expressed the scale factor and the scalar field in closed-form functions.

We have found five families of scalar field potentials which are Liouville--integrable, 
and that admits conservation laws quadratic  in the momenta, and they are in the form
$V_{A}\left(  \phi\right)  =V_{0}\phi^{p}+V_{1}\phi^{r}$, where $p,r$ are
constants. For each dominant term of the potential the analytic solution for
the scale factor behaves like a power-law function or like an exponential function
which describes the de Sitter universe when the dominant power has the value
two. We remark that we have selected a specific interaction function between
the aether and the scalar fields. The interaction form that we selected has
also geometric origins since for that function in the minisuperspace
description, the dynamical variables of the field equations evolve in a
two-dimensional space of maximally symmetry; in particular in two-dimensional
flat space of Lorentzian signature. That is also a condition that we have
assumed, in order the field equations to admit conservation laws quadratic in
the momentum.

For some of the close-form solutions that we have found, we have studied the 
qualitative behavior of the Hubble factor, and we have presented the evolution 
of the equation of state parameter of the effective fluid in terms of redshift. From 
which we found that the effective fluid can cross the phantom divide line and
behaves like a quintom field \cite{Dutta:2009yb,Guo:2004fq,Zhao:2006mp,Lazkoz:2006pa,Lazkoz:2007mx,MohseniSadjadi:2006hb,Setare:2008pz,Setare:2008dw,Saridakis:2009ej,Cai:2009zp,Qiu:2010ux,Leon:2012vt,Leon:2018lnd,Paliathanasis:2018vru} or as a phantom field \cite{Singh:2003vx,Sami:2003xv,Andrianov:2005tm,Elizalde:2008yf,Sadatian:2008sv}. However the final attractor for that solutions is that of the de Sitter universe. More analysis should be done in that models in
order to specify their physical viability, specifically if they can be contrasted against of
cosmological observations. However such an analysis extends the scopes of this
work and will be published elsewhere.

\begin{acknowledgments}
G. L. was funded by Agencia Nacional de Investigaci\'{o}n y Desarrollo--ANID
through the program FONDECYT Iniciaci\'{o}n grant no. 11180126. 
Additionally, G. L. acknowledges the financial support of Vicerrector\'{\i}a de Investigaci\'{o}n y Desarrollo Tecnol\'{o}gico at
Universidad Catolica del Norte.
\end{acknowledgments}

\appendix

\section{Liouville integrable potentials with arbitrary parameter $\gamma$}
\label{appa}
In the previous analysis we presented the potentials $V\left(\phi\right)$, leading to Liouville--integrable gravitational field equations with a dust fluid source. In this Appendix, we present the potentials where the field equations are Liouville--integrable for arbitrary values of the barotropic index $\gamma$ of the ideal gas. In this regard, the minisuperspace Lagrangian reads
\begin{equation}
L\left(  N,a,\dot{a},\phi,\dot{\phi}\right)  =a^{3\left(  \gamma-1\right)
}\left(  -3B\left(  \phi\right)  a\dot{a}^{2}+\frac{1}{2}a^{3}\dot{\phi}%
^{2}\right)  -a^{-3\left(  \gamma-2\right)  }V\left(  \phi\right)  -\rho_{m0}.
\label{app.01}%
\end{equation}
where have considered $N\left(  t\right)  =a^{-3\left(  \gamma-1\right)}.$

Hence, for Lagrangian (\ref{app.01}) we find five potentials which are
Liouville integrable, of the form%
\begin{equation}
V_{LI}\left(  \phi\right)  =V_{0}\phi^{P_{0}}+V_{1}\phi^{P_{1}},
\end{equation}
where $\left\{  P_{0},P_{1}\right\}$ take values on the following sets:%
\begin{equation}
\bar{V}_{A}\left(  \phi\right)  \text{ for }\left\{  P_{0},P_{1}\right\}   
=\left\{  2-\frac{\gamma-1}{\sqrt{B_{0}}},-\frac{\gamma-2}{2\sqrt{B_{0}}%
}\right\},
\end{equation}
\begin{equation}
\bar{V}_{B}\left(  \phi\right)  \text{ for }\left\{  P_{0},P_{1}\right\}   
=\left\{  2+\frac{\gamma-1}{\sqrt{B_{0}}},\frac{\gamma-2}{2\sqrt{B_{0}}%
}\right\},
\end{equation}
\begin{equation}
\bar{V}_{C}\left(  \phi\right)  \text{ for }\left\{  P_{0},P_{1}\right\}   
=\left\{  -\frac{2}{\gamma}\left(  \gamma-2\right)  ,-2+\frac{\gamma}{4B_{0}%
}\right\},
\end{equation}
\begin{equation}
\bar{V}_{D}\left(  \phi\right)  \text{ for }\left\{  P_{0},P_{1}\right\}   
=\left\{  \frac{1}{2\sqrt{A _{0}}}-\frac{\gamma-1}{2\sqrt{B_{0}}}%
,-1-\frac{\left(  \gamma-4\right)  \left(  \gamma+4\sqrt{B_{0}}\right)
}{4\left(  4B_{0}+\gamma\sqrt{B_{0}}\right)  }\right\},
\end{equation}
\begin{equation}
\bar{V}_{E}\left(  \phi\right)  \text{ for }\left\{  P_{0},P_{1}\right\}   
=\left\{  \frac{\left(  2-\gamma\right)  \left(  \gamma-4\sqrt{B_{0}}\right)
}{2\left(  4B_{0}-\sqrt{B_{0}}\gamma\right)  },-1+\frac{\left(  \gamma
-4\right)  \left(  4\sqrt{B_{0}}-\gamma\right)  }{4\left(  4B_{0}-\gamma
\sqrt{B_{0}}\right)  }\right\}.
\end{equation}
The transformation of the canonical variables in this model now is defined as%
\begin{equation}
\bar{x}=a^{6\sqrt{B_{0}}+\frac{3}{2}\gamma}\phi^{1+\frac{\gamma}{4\sqrt{B_{0}%
}}}~,~\bar{y}=a^{-6\sqrt{B_{0}}+\frac{3}{2}\gamma}\phi^{1-\frac{\gamma}%
{4\sqrt{B_{0}}}}.
\end{equation}

As this point it is important to mention that the above results include the
case where the Einstein-aether scalar field theory it is on a FLRW spacetime
with nonzero spatially curvature $K$, that it is true when $\gamma=\frac{2}%
{3}$ and $\rho_{m0}=K$.

\end{document}